\documentclass[prd,nofootinbib,superscriptaddress,twocolumn,floatfix]{revtex4}
\usepackage{times} %
\usepackage{pdfsync}%

\usepackage{amssymb,amsfonts,amsmath,paralist,xspace} %

\usepackage{longtable}%
\usepackage[normalem]{ulem} %
\usepackage{relsize,color,xcolor} %
\usepackage{booktabs} 

\usepackage{units} 

\usepackage{grffile} %
\usepackage{graphicx} %
\graphicspath{{./figures/}} 

\IfFileExists{srcltx.sty}{\usepackage[active]{srcltx}}

\usepackage{etoolbox} %
\preto\tabular{\setcounter{magicrownumbers}{0}}%
\newcounter{magicrownumbers}%
\newcommand\rownumber{\stepcounter{magicrownumbers}\arabic{magicrownumbers}}%

\newcommand{\keV}{\:\mathrm{keV}} 

 \newcommand{\be}
{\begin{equation}} \newcommand{\ee} {\end{equation}}   

\def\be{\begin{eqnarray}} \def\ee{\end{eqnarray}} 
 \renewcommand{\S}{\mathcal{S}} %
\newcommand{\dm}{{\textsc{dm}}} 
\newcommand{\xmm}{\textit{XMM-Newton}\xspace}
\newcommand{\fov}{\mathrm{fov}} 
 
\renewcommand{\S}{\mathcal{S}} 

\begin{document}

\title{An unidentified line in X-ray spectra of the Andromeda galaxy and
  Perseus galaxy cluster}

\author{A.~Boyarsky$^{1}$, O.~Ruchayskiy$^{2}$, D.~Iakubovskyi$^{3,4}$ and J.~Franse$^{1,5}$\\
  $^1${\small Instituut-Lorentz for Theoretical Physics, Universiteit Leiden, Niels Bohrweg 2, Leiden, The Netherlands}\\
  $^2${\small Ecole Polytechnique F\'ed\'erale de Lausanne, FSB/ITP/LPPC, BSP, CH-1015, Lausanne, Switzerland}\\
  $^3${\small Bogolyubov Institute of Theoretical Physics, Metrologichna Str. 14-b, 03680, Kyiv, Ukraine}\\
  $^4${\small National University ``Kyiv-Mohyla Academy'', Skovorody Str. 2, 04070, Kyiv, Ukraine}\\
  $^5${\small Leiden Observatory, Leiden University, Niels Bohrweg 2, Leiden, The Netherlands}\\
}

\begin{abstract}
  We report a weak line at $3.52\pm 0.02$~keV in X-ray spectra of M31 galaxy and
  the Perseus galaxy cluster observed by MOS and PN cameras of XMM-Newton
  telescope.  This line is not known as an atomic line in the spectra of
  galaxies or clusters. It becomes stronger towards the centers of the
  objects; is stronger for Perseus than for M31; is absent in the spectrum of
  a deep ``blank sky'' dataset.  Although for each object it is hard to exclude
  that the feature is due to an instrumental effect or an atomic line, it is
  consistent with the behavior of a dark matter decay line.  Future
  (non-)detections of this line in multiple objects may help to reveal its
  nature.

\end{abstract}

\maketitle

The nature of dark matter (DM) is a question of crucial importance for both
cosmology and for fundamental physics. As neutrinos -- the only known
particles that could be DM candidates -- are too light to be consistent with
various observations~\cite{Tremaine:79,White:83,Hannestad:03,Boyarsky:08a}, it
is widely anticipated that new particles should exist. Although many
candidates have been put forward (see e.g.~\cite{Feng:10}), little is known
experimentally about the properties of DM particles: their masses, lifetimes,
and interaction types remain largely unconstrained.  \textit{A priori}, a
given DM candidate can possess a decay channel if its lifetime exceeds the age
of the Universe. Therefore, the search for a DM decay signal provides an
important test to constrain the properties of DM in a model-independent
way. For fermionic particles, one should search above the Tremaine-Gunn
limit~\cite{Tremaine:79} ($\gtrsim \unit[300]{eV}$). If the mass is below $2
m_e c^2$, such a fermion can decay to neutrinos and photons with energy
$E_\gamma=\frac12 m_\dm$~\cite{Pal:82}. One can search for such particles in
X-rays~\cite{Abazajian:01b,Dolgov:00} (see~\cite{Boyarsky:12c} for review of
previous searches).  For each particular model, the particle's parameters are
related by the requirement to provide the correct DM abundance.  For example,
for one very interesting DM candidate -- the right-handed neutrino -- this
requirement restricts the mass range to
$\unit[0.5-100]{keV}$~\cite{Boyarsky:09a,Boyarsky:12c}.  A large part of the
available parameter space for sterile neutrinos is consistent with all
astrophysical and cosmological bounds~\cite{Boyarsky:08d}, and it is important
to probe it further.

The DM decay line is much narrower than the spectral resolution of the existing X-ray telescopes 
and, as previous searches have shown, should be rather
weak. The X-ray spectra of astrophysical objects are crowded with weak atomic
and instrumental lines, not all of which may be known.  Therefore, even if the
exposure of available observations continues to increase, it is hard to
exclude an astrophysical or instrumental origin of any weak line found in the
spectrum of individual object. However, if the same feature is present in the
spectra of many different objects, and its surface brightness and
relative normalization between objects are consistent with the expected
behavior of the DM signal, this can provide much more convincing evidence
about its nature.

The present paper takes a step in this direction.  We present the results of
the combined analysis of many \xmm\ observations of two objects at different
redshifts -- the Perseus cluster ($z = 0.0179$~\cite{NED}) and the Andromeda
galaxy (M31, $z = -0.001$)\footnote{Each of the datasets used in previous
  decaying DM searches in
  M31~\cite{Watson:06,Boyarsky:07a,Boyarsky:10b,Watson:11,Horiuchi:13} had
  less statistics than we use. The non-detection of any signal in these works
  does not come in contradiction with our results.}, a Local Group member --
together with a long exposure ``blank sky'' dataset.
We present the detection of a significant un-modeled excess at $3.52\pm 0.02$~keV (restframe) in both objects. 
We study its behaviour and establish that it is consistent with a DM interpretation. 
However, as the
line is weak ($\sim 4\sigma$ in the combined dataset) and the uncertainties in
DM distribution are significant, positive detections or strong constraints
from more objects are clearly needed
to determine 
its nature.\footnote{When this paper was in preparation, the arXiv preprint of~\cite{Bulbul:14}
  appeared, claiming a detection of a spectral feature at the same energy from
  a collection of galaxy clusters. Our analyses are independent, based on
  different datasets, but the results are fully consistent.}

Below we summarize the details of our data analysis and then discuss the results and caveats.

\textbf{Data analysis.} 
We use the data obtained with MOS~\cite{Turner:00} and PN~\cite{Strueder:01}
CCD cameras of \xmm (``XMM'' in what follows).  We use SAS
v.13.0.0~\cite{XMM-SAS} to reduce the raw data and filter the data for
\emph{soft solar protons}~\cite{Read:03,Kuntz:08} using the \texttt{espfilt}
procedure.
Because residual soft proton flares can produce weak line-like features in the
spectra at positions where the effective area is non-monotonic (see
e.g.~\cite{Boyarsky:10a}), we apply the procedure described
in~\cite{DeLuca:03}, based on the comparison of high-energy count rates for
``in-FoV'' (10-15~arcmin off-center) and out-FoV CCD
regions~\cite{Fin_over_Fout}.  We selected only observations where the ratio
of $F_{in}-F_{out} < 1.15$.\footnote{Ref.~\cite{DeLuca:03} argued that
  $F_{in}-F_{out} < 1.3$ is a sufficient criterion for flare removal. We find
  by visual inspection of the resulting spectra that a stricter criterion is
  needed to reduce artificial line-like
  residuals~\cite{Boyarsky:10a,DimaPhD}. Lowering the threshold further is not
  feasible as the statistical errorbars on the value of $F_{in}-F_{out}$ are
  of the order of $5\%$. }

\textbf{Combined observation of M31.}  We use $\sim 2$~Msec of raw
exposure observations of M31 within the central $1.5^\circ$ (see SOM, Table~II).
We select from the XMM archive 29 MOS observations offset less than $1.5'$ from the center of M31, 
and 20 MOS observations with offsets $23.7' - 55.8'$ that passed our criterion for residual contamination.
Not enough PN observations passed this test to include them. 
The central and off-center observations were co-added seperately with the \texttt{addspec} routine from 
\textsc{ftools}~\cite{ftools}.
The resulting spectra were binned by 60~eV.
This bin size is a factor $\sim 2$ smaller than the spectral resolution of the XMM at
these energies, which makes the bins roughly statistically independent.

\begin{table*}[t!]
  \centering
  \begin{tabular}[c]{l|c|c|c|c|c|c}
    Dataset & Exposure [ksec]& $\chi^2/\text{d.o.f.}$ & Line position [keV] & Flux [$\unit[10^{-6}]{cts/sec/cm^2}$]&
    $\Delta \chi^2$ & Significance \\
    \hline
    {\sc M31 on-center} & 978.9 & 97.8/74 & $3.53 \pm 0.03$ &
    $4.9_{-1.3}^{+1.6}$ & 13.0 & $3.2\sigma$\\
    \hline
    {\sc M31 off-center} & 1472.8 & 107.8/75 &  $3.50 - 3.56$ &  $<1.8$ ($2\sigma$) & \dots\\
    \hline
    {\sc Perseus cluster (MOS) } & 628.5 & 72.7/68  & $ 3.50\pm 0.04$ &
    $7.0^{+2.6}_{-2.6}$ & 9.1 & $2.6\sigma$\\
    {\sc Perseus cluster (PN)} & 215.5 & 62.6/62  & $ 3.46\pm 0.04$ &
    $9.2^{+3.1}_{-3.1}$ & 8.0 & $2.4\sigma$\\
    \hline
    {\sc Perseus (MOS)} & 1507.4 & 191.5/142 & $3.52\pm 0.02$ &
    $8.6^{+2.2}_{-2.3}$  ({\footnotesize Perseus})& 25.9 & $4.4\sigma$\\
    {\sc   + M31 on-center} & &           &                        & $4.6^{+1.4}_{-1.4}$
    ({\footnotesize M31})  & (3 dof)\\
    \hline 
    {\sc Blank-sky}& 15700.2 & 33.1/33 & $3.45 - 3.58$ & $<0.7$ ($2\sigma$) & \dots\\
    \hline 
  \end{tabular}
  \caption{\label{tab:m31_chi2} Basic properties of combined observations used in this paper. 
    Second column denotes the sum of exposures of individual observations. 
    The improvement in
    $\Delta \chi^2$ when extra line is added to a model is quoted for each
    dataset. The last column shows the
    local significance of such an improvement when 2 extra d.o.f.\ (position
    and flux of the line) are  added. 
    The energies for Perseus are quoted in the rest frame. Taking into account
    trial factors, the global (over
    three datasets) significance is $4.4\sigma$ (see SOM for details).} 
\end{table*}

\textbf{Background modeling.}  We model the contribution of the instrumental
(particle induced) background by a combination of an unfolded power law
plus several narrow
\texttt{gaussian} lines. The positions and normalizations of the lines
were allowed to vary freely and the most prominent instrumental K-$\alpha$
lines (Cr, Mn, K, Fe, Ni, Ca, Cu) and Fe K$\beta$ have been recovered.  The
width of the Gaussians was fixed at 1~eV (an infinitely thin line for the XMM
spectral resolution). We verified that allowing the line widths to vary freely
leaves the results unchanged. We restrict our modeling to the energy interval
2--8~keV. The Galactic foreground is negligible above
2~keV~\cite{Nevalainen:05}. The combined emission of unresolved point sources
at these energies is modeled by a
\texttt{powerlaw}~\cite{Takahashi:04}. Several line-like residuals around
2.4~keV and 3.0~keV were identified as Ar and S line complexes and the
corresponding thin (1 eV width) lines were added to the model. 
We verified that adding another
\texttt{powerlaw} component to model the contribution of the extragalactic
X-ray background~\cite{DeLuca:03,Nevalainen:05} does not improve the quality
of fit and does not change the structure of the residuals.

 \begin{figure*}[tp!]
  \centering

  \includegraphics[width=0.45\textwidth]{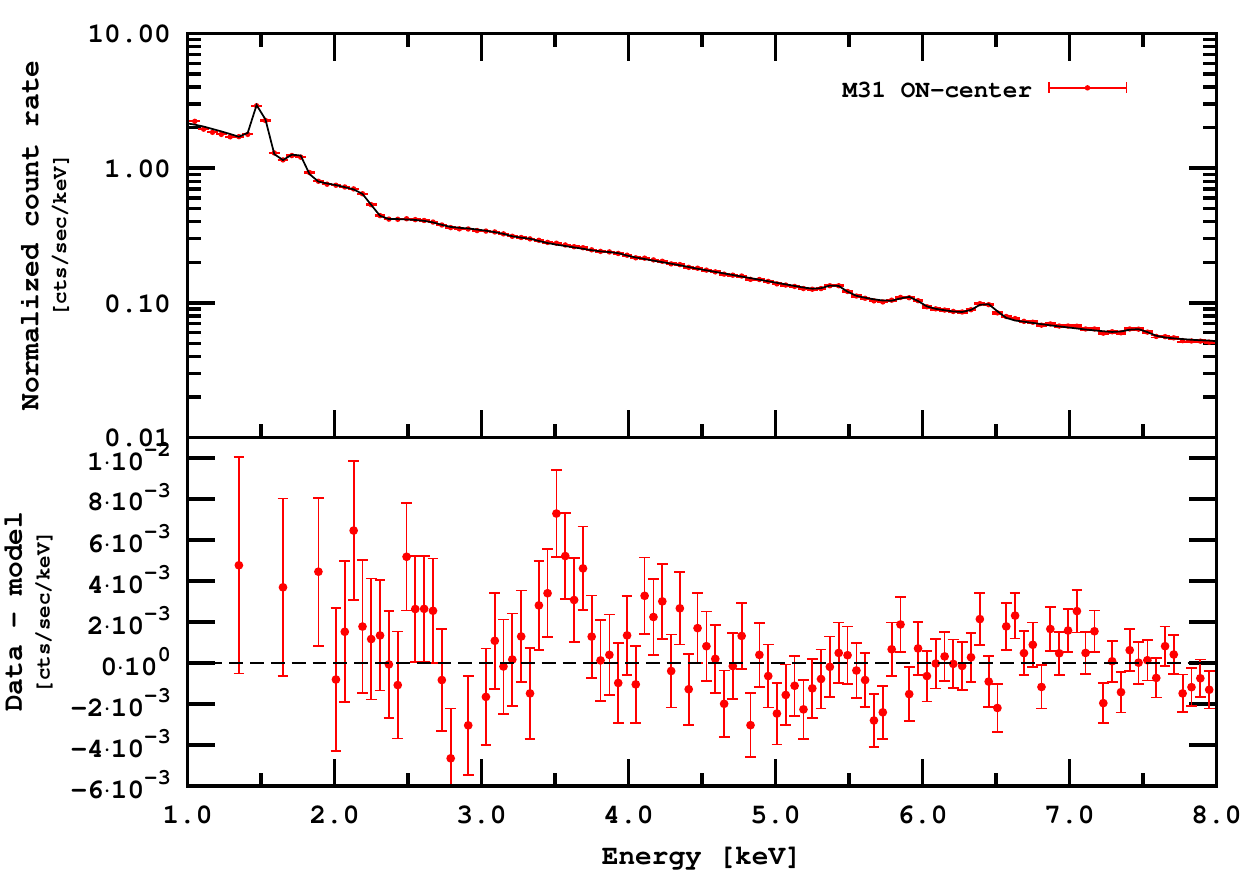}~%
  \includegraphics[width=0.45\textwidth]{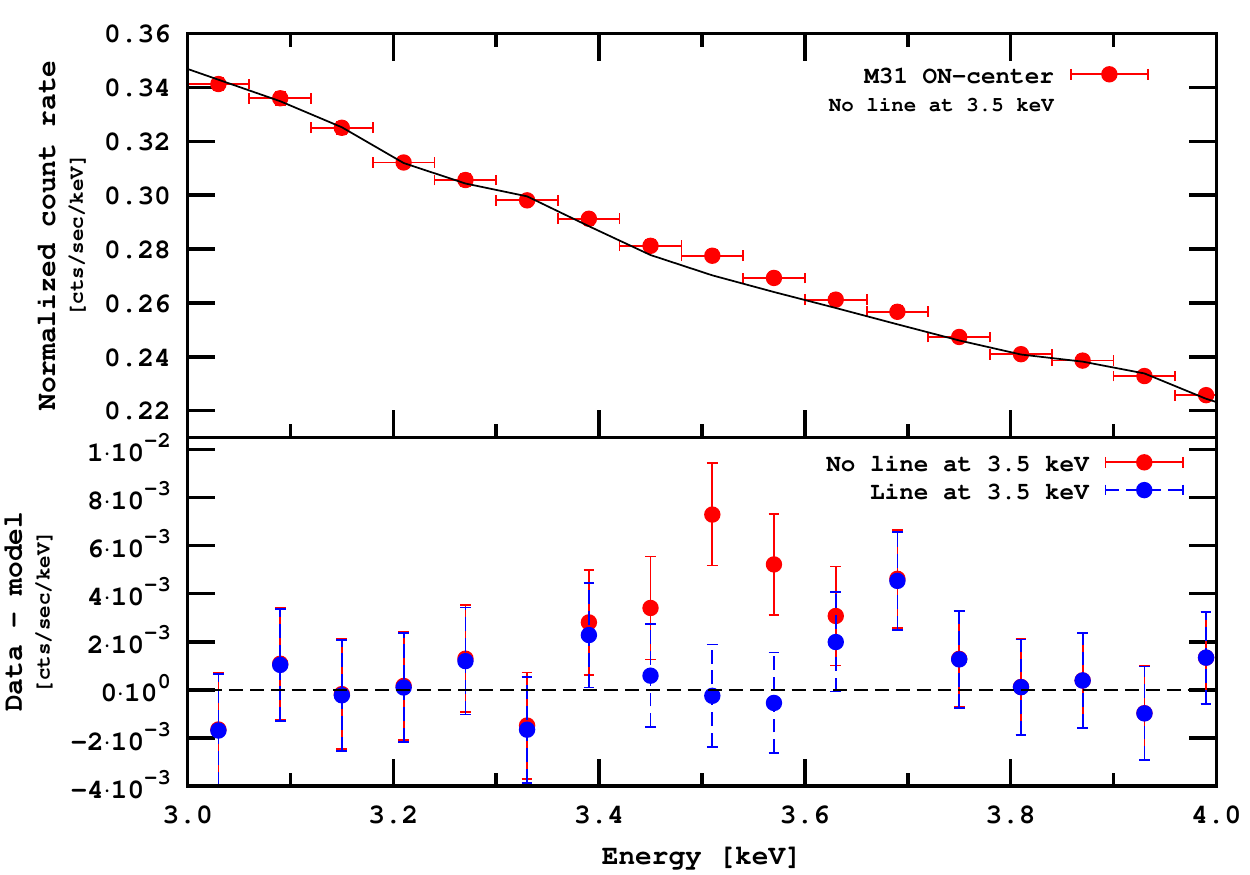}
  \caption{\textit{Left:} Folded count rate (top) and residuals (bottom) for
    the MOS spectrum of the central region of M31. Statistical Y-errorbars on the top
    plot are smaller than the point size. The line around 3.5~keV is \emph{not
      added}, hence the group of positive residuals. \textit{Right}: zoom onto
    the line region.}
  \label{fig:m31_oncen}
\end{figure*}

\textbf{Result.}  The resulting spectrum of the central observations shows a group of positive residuals
around $3.5$~keV (Fig.~\ref{fig:m31_oncen}). Adding a thin Gaussian line at
that energy reduces the total $\chi^2$ by $\sim 13$, see
Table~\ref{tab:m31_chi2} (more than $3\sigma$ significance for extra 2 degrees
of freedom). Examination of MOS1 and MOS2 observations individually finds the
line in both cameras with comparable flux.
For the off-center observations, none of the cameras show any detectable residual in the energy range
$3.50-3.56$~keV. The $2\sigma$ upper bound on the flux is given in
Table~\ref{tab:m31_chi2}.

\textbf{Perseus cluster.} If the candidate weak signal is of astrophysical
(rather than instrumental) origin, we should be able to detect its
redshift. To this end we have chosen the nearby Perseus cluster (Abell 426).
At its redshift the line's centroid would be shifted by 63~eV.  As the
position of the line is determined with about 30~eV precision, one can expect
to resolve the line's shift with about $2\sigma$ significance.

We took 16 off-center observations of the Perseus cluster (SOM, Table~II) and
processed them in the same way as for M31.  The flare removal procedure left
215~ksec of PN camera's exposure, therefore we also use PN data.

\textbf{Background modeling.} The resulting spectra were then added together
and fitted to the combination of \texttt{vmekal} (with free abundances for Fe,
Ni, Ar, Ca and S) plus (extragalactic) \texttt{powerlaw}. The instrumental
background was modeled as in the M31 case. 

\textbf{Results.} The fit shows significant positive residuals at energies
around 3.47~keV (in the detector frame). Adding a \texttt{zgauss} model with
the redshift of the cluster improves the fit by $\Delta\chi^2 = 9.1$. The
line's position is fully consistent with that of M31
(Table~\ref{tab:m31_chi2}). If we fix the position of the line to that of M31
and allow the redshift to vary, $z=0$ provides a worse fit by $\Delta \chi^2 =
3.6$ and its best-fit value is $(1.73\pm 0.08) \times 10^{-2}$ -- close to the
value $z=0.0179$~ which we have used.

\textbf{Blank-sky dataset}. To further study the origin of the new line and
possible systematic effects we combine XMM blank-sky observations
from~\cite{Carter:07,Henley:12} with observations of the Lockman
Hole~\cite{Brunner:07}. The data were reduced similarly to the other datasets.
Fig.~\ref{fig:blanksky} shows the combined spectrum.  A dataset with such a
large exposure requires special analysis (as described
in~\cite{DimaPhD}). This analysis did not reveal any line-like residuals in
the range $3.45 - 3.58$~keV with the $2\sigma$ upper bound on the flux being
$\unit[7\times 10^{-7}]{cts/cm^2/sec}$. The closest detected line-like feature
($\Delta\chi^2 = 4.5$) is at $3.67^{+0.10}_{-0.05}$~keV, consistent with the
instrumental Ca~K$\alpha$ line.\footnote{Previously this line has only been
  observed in the PN camera~\cite{Strueder:01}.}

\textbf{Combined fit of M31 + Perseus.} Finally, we have performed a
simultaneous fit of the on-center M31 and Perseus datasets (MOS), keeping a
common position of the line (in the rest-frame) and allowing the line
normalizations to be different. The line improves the fit by $\Delta \chi^2 =
25.9$ -- $4.4\sigma$ significance (Table~\ref{tab:m31_chi2}).

\textbf{Results and discussion.} We identified a spectral feature at
$E=3.52\pm 0.02$~keV in the combined dataset of M31 and Perseus with a
statistical significance $4.4\sigma$ which does not coincide with any known
line. Next we compare its properties with the expected behavior of a DM decay
line.

The observed brightness of a decaying DM should be proportional to its column density 
$\S_\dm=\int \rho_\dm d\ell$ -- integral along the line
of sight of the DM density distribution -- and inversely proportional to the
radiative decay lifetime $\tau_\dm$:
\begin{eqnarray}
  && F_\dm\approx 2.0\times 10^{-6} \frac{\unit{cts}}{\unit{cm^2\cdot sec}} 
  \left( \frac{\Omega_\fov}{500~\mbox{arcmin}^2}\right) \times \\ \nonumber
  && \left(\frac{\S_\dm}{500~\unit{M_{\bigodot}/{pc}^2}}\right)
  \frac{10^{29}~\unit{s}}{\tau_\dm}\left(\frac{\unit{keV}}{m_\dm}\right).
\label{eq:DM_flux}
\end{eqnarray}

\begin{figure*}[!t]
  \centering
  \includegraphics[height=0.45\linewidth,angle=-90]{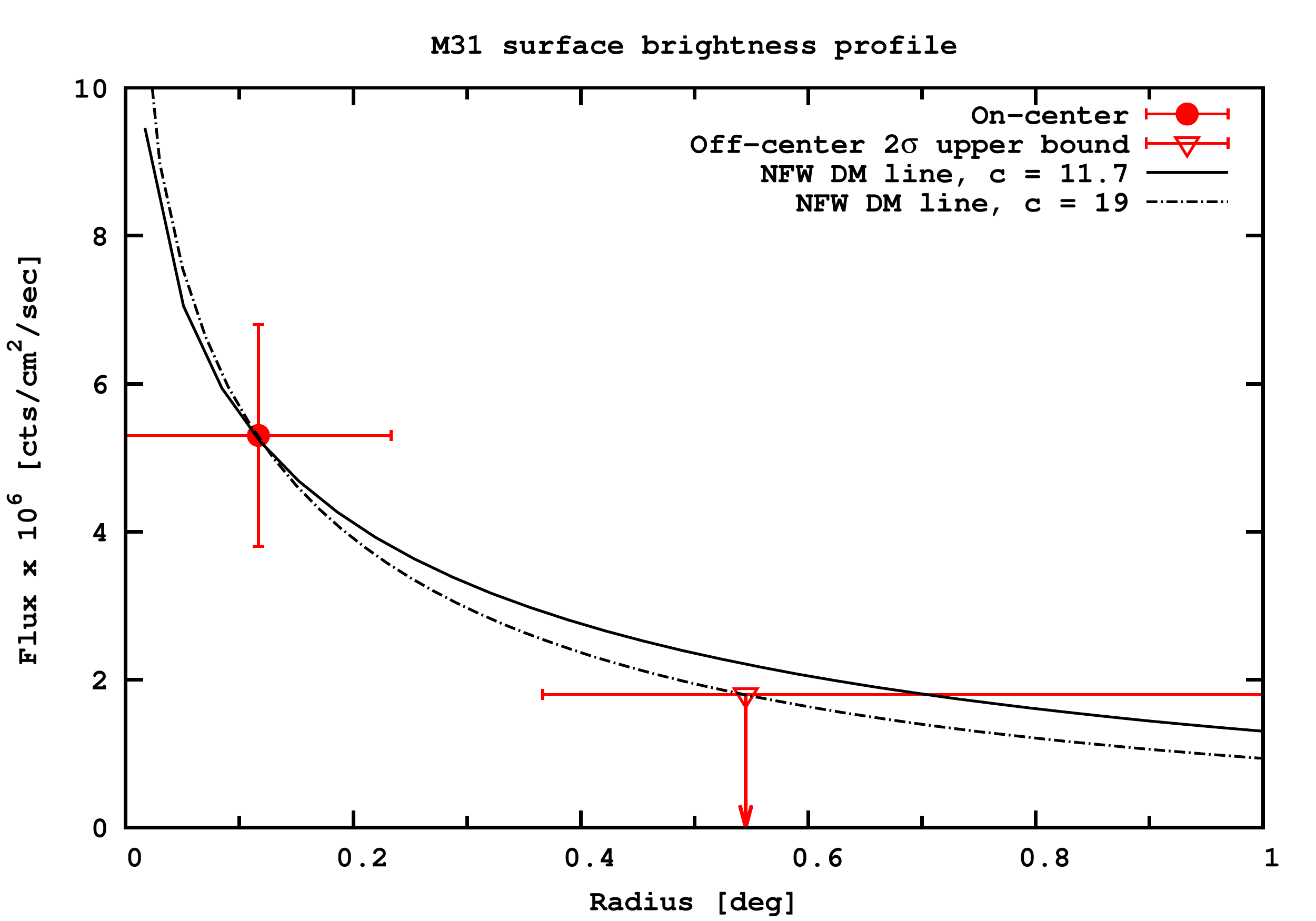}
  ~\includegraphics[height=0.45\linewidth,angle=-90]{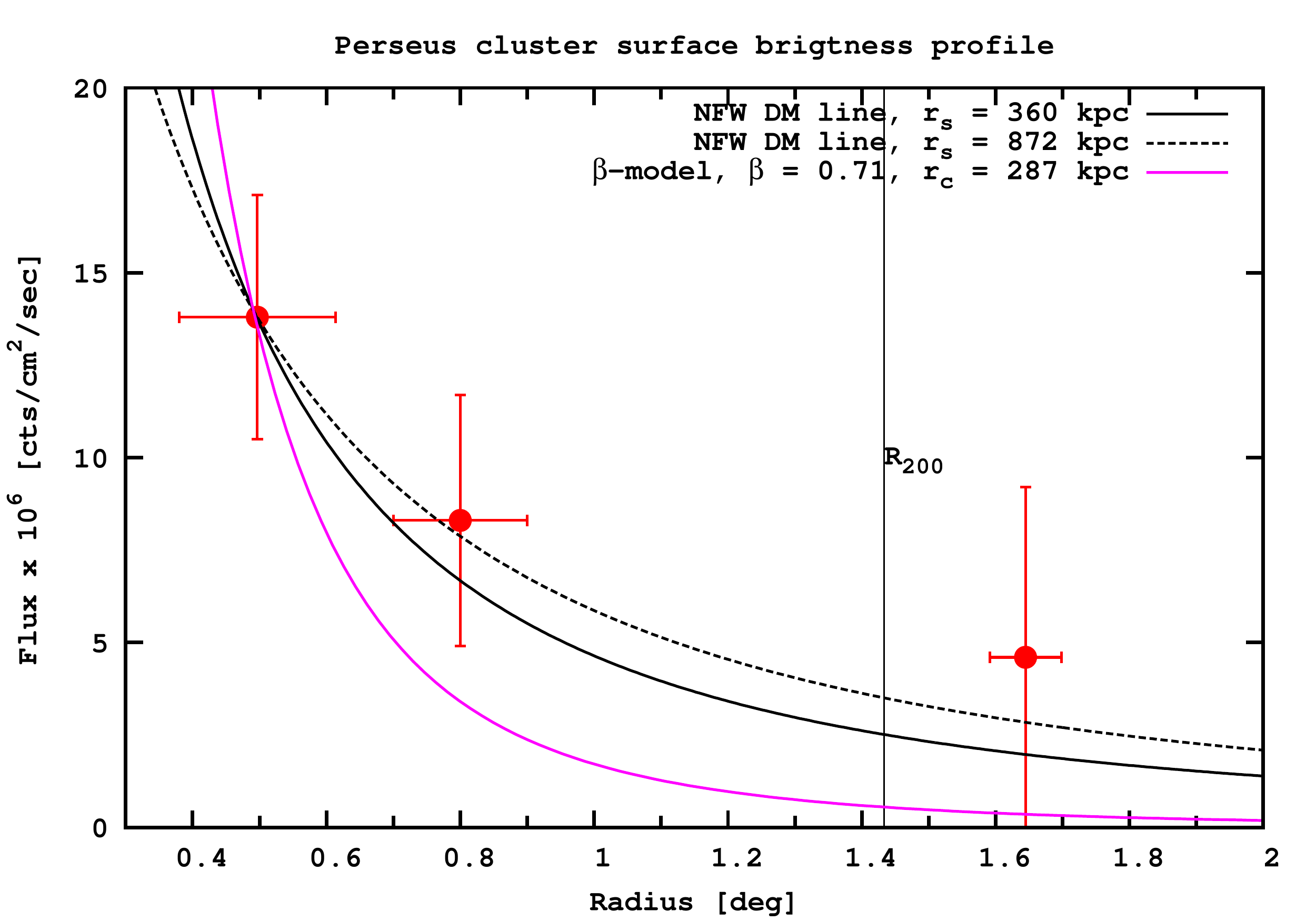}
\caption{The line's brightness profile in M31 (left) and the Perseus cluster
  (right). A NFW DM distribution is assumed, the scale $r_s$ is fixed to its
  best-fit values from~\protect\cite{Corbelli:09} (M31) or
  \protect\cite{Simionescu:11} (Perseus) and the overall normalization is
  adjusted to pass through the left-most point.}
  \label{fig:radial_profile}
\end{figure*}
\textbf{M31 and Perseus brightness profiles.} Using the line flux of the
center of M31 and the upper limit from the off-center observations we
constrain the spatial profile of the line.  The DM distribution in M31 has
been extensively studied (see an overview in~\cite{Boyarsky:10a}).  We take
NFW profiles for M31 with concentrations $c=11.7$ (solid line,
\cite{Corbelli:09}) and $c=19$ (dash-dotted line). For each concentration we
adjust the normalization so it passes through first data point
(Fig.~\ref{fig:radial_profile}).  The $c= 19$ profile was chosen to intersect
the upper limit, illustrating that the obtained line fluxes of M31 are fully
consistent with the density profile of M31 (see
e.g.~\cite{Corbelli:09,Chemin:09,SanchezConde:2011ap} for a $c = 19-22$ model
of M31).

For the Perseus cluster the observations can be grouped in 3 radial bins by
their off-center angle. For each bin we fix the line position to its average
value across Perseus ($3.47\pm 0.07$~keV). The obtained line fluxes together
with 1$\sigma$ errors are shown in Fig.~\ref{fig:radial_profile}. For
comparison, we draw the expected line distribution from DM decay
using the NFW profile of~\cite{Simionescu:11} (best fit value $r_s = 360$~kpc
($c\approx5$), black solid line;  upper bound $r_s = 872$~kpc
($c\approx 2$), black dashed line). The isothermal $\beta$-profile
from~\cite{Urban:13} is shown in magenta.  The surface brightness profile
follows the expected DM decay line's distribution in Perseus.

Finally, we compare the predictions for the DM lifetime from the two
objects. The estimated column density within the central part of M31 ranges
between $\bar{\mathcal{S}} \sim \unit[200 - 1000]{M_\odot/pc^2}$ with the
average value being around $\unit[600]{M_\odot/pc^2}$~\cite{Boyarsky:10a}. The
column density of clusters follows from the $c{-}M$
relation~\cite{Boyarsky:09c,King:11,Mandelbaum:08}. Considering the
uncertainty on the profile and that our observations of Perseus go beyond
$r_s$, the column density in the region of interest is within
$\bar{\mathcal{S}} \sim \unit[100 -600]{M_\odot/pc^2}$.  Therefore the ratio
of expected signals between Perseus and the center of M31 can be $0.1 - 3.0$,
consistent with the ratio of measured fluxes $0.7 - 2.7$.

If DM is made of right-handed (sterile) neutrinos~\cite{Dodelson:93}, the
lifetime is related to its interaction strength (\emph{mixing angle}):
\begin{small}
  \begin{displaymath} \tau_\dm = \frac{1024\pi^{4}}{9 \alpha
      G_{F}^{2}\sin^{2}( 2\theta) m_{\dm}^{5}} = 7.2 \times 10^{29} \sec
    \left[\frac{10^{-8}}{\sin^{2}(2\theta)}\right]\left[\frac{1\keV}
      {m_\dm}\right]^{5}.
  \end{displaymath}
\end{small}

Using the data from M31 and taking into account uncertainties in its DM
content we obtain the mass $m_\dm = 7.06 \pm 0.06$~keV and the mixing angle in
the range $\sin^2(2\theta) = (2 - 20)\times 10^{-11}$ (taking the column
density $\bar S = \unit[600]{M_\odot/pc^2}$ and using only statistical
uncertainties on flux we would get $\sin^2(2\theta) = 4.9^{+1.6}_{-1.3}\times
10^{-11}$ ).  This value is fully consistent with previous bounds,
Fig.~\ref{fig:numsm}.  Moreover, it is intriguing that this value is
consistent with the result of the paper~\cite{Bulbul:14}, which appeared when
our paper was in preparation. Indeed, our value of $\sin^2(2\theta)$ is based
on completely independent analysis of the signal from M31 and our estimates
for its DM content, whereas the result of~\cite{Bulbul:14} is based on the
signal from stacked galaxy clusters and on the weighted DM column density from
the full sample.

These values of $\sin^2(2\theta)$ means that sterile neutrinos should be
produced resonantly~\cite{Shi:98,Shaposhnikov:08a,Laine:08a}, which requires
the presence of significant lepton asymmetry in primordial plasma at
temperatures few hundreds MeV. This produces restrictions on parameters of the
$\nu$MSM~\cite{Boyarsky:09a}.

\begin{figure}[!tp]
  \centering
  \includegraphics[width=\linewidth]{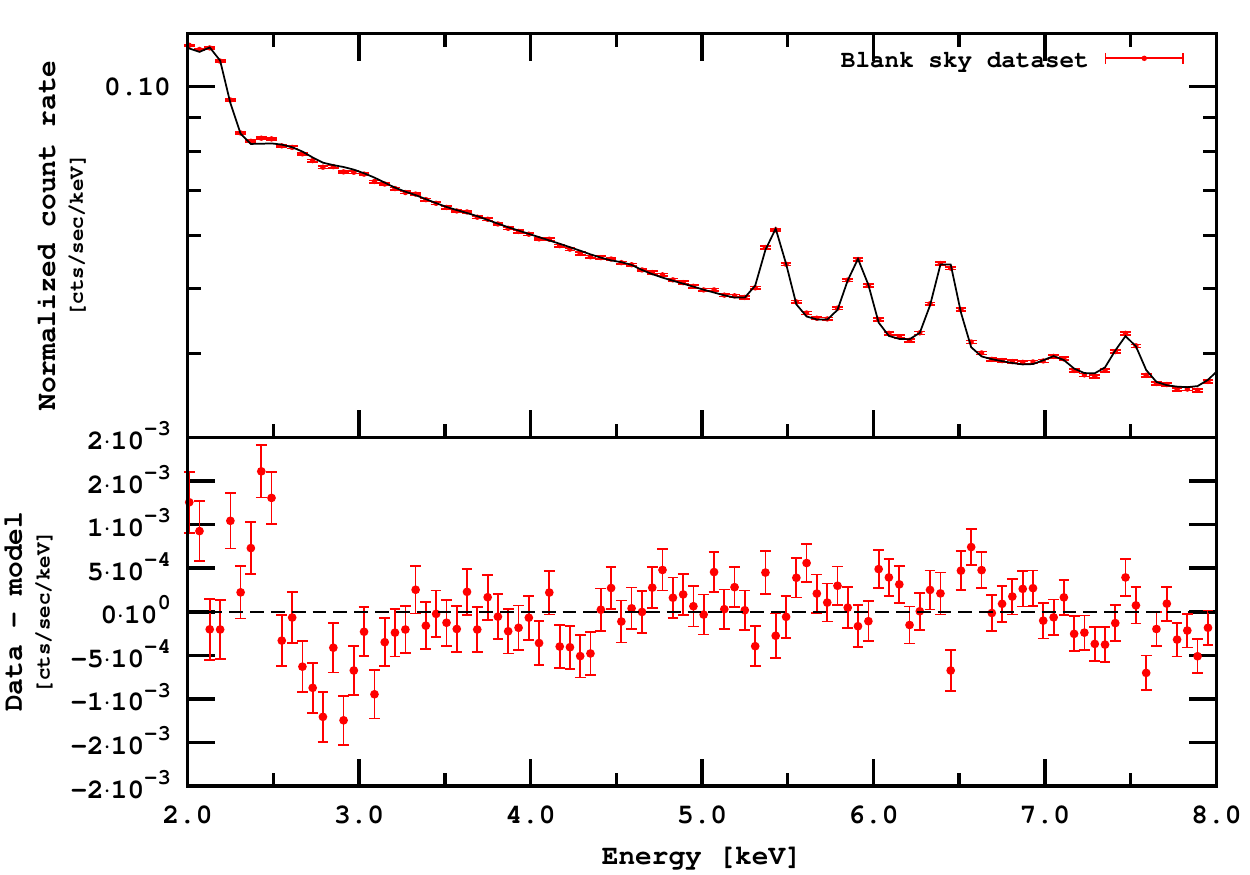}
  \caption{Combination of 382 MOS blank sky observations.}
  \label{fig:blanksky}
\end{figure}

The position and flux of the discussed weak line are inevitably subject to
systematical uncertainties. There are two weak instrumental lines (K~K$\alpha$
at 3.31~keV and Ca~K$\alpha$ at 3.69~keV), although formally their centroids
are separated by more than $4\sigma$. Additionally, the region below $3$~keV
is difficult to model precisely, especially at large exposures, due to the
presence of the absorption edge and galactic emission.  However, although the
residuals below 3~keV are similar between the M31 dataset
(Fig.~\ref{fig:m31_oncen}) and the blank sky dataset
(Fig.~\ref{fig:blanksky}), the line is \emph{not detected} in the latter.

If the feature were due to an unmodelled wiggle in the effective area, its
flux would be proportional to the continuum brightness and the blank-sky
dataset would have exhibited a 4 times smaller feature with roughly the same
significance (see SOM, Section B). In addition, the Perseus line would not be
properly redshifted.

The properties of this line are consistent (within
uncertainties) with the DM interpretation.  To reach a conclusion about its
nature, one will need to find more objects that give a detection or where
non-observation of the line will put tight constraints on its properties.  The
forthcoming \textit{Astro-H} mission~\cite{Takahashi:12} has sufficient
spectral resolution to spectrally resolve the line against other nearby
features and to detect the candidate line in the ``strong line''
regime~\cite{Boyarsky:06f}.  In particular, \textit{Astro-H} should be able to
resolve the Milky Way halo's DM decay signal and therefore all its
observations can be used. Failure to detect such a line will rule out the DM
origin of the Andromeda/Perseus signal presented here.
\begin{figure}[!t]
  \centering
  \includegraphics[width=\linewidth]
  {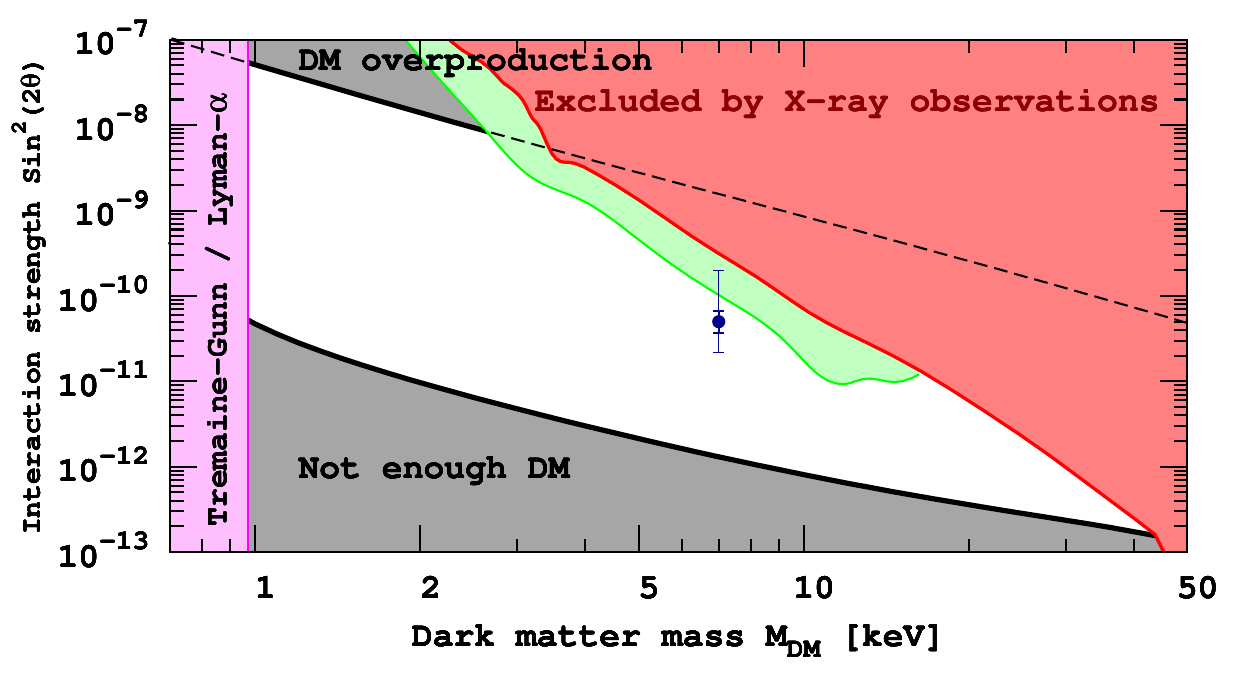}
  \caption{Constraints on sterile neutrino DM within
    $\nu$MSM~\cite{Boyarsky:12c}. Recent bounds
    from~\cite{Watson:11,Horiuchi:13} are shown in green. Similar to older
    bounds (marked by red) they are smoothed and divided by factor 2 to
    account for possible DM uncertainties in M31.
    In every point in the white region sterile
    neutrino constitute 100\% of DM and their properties agree with
    the existing bounds. Within the gray regions too much (or not enough) DM
    would be produced in a minimal model like $\nu$MSM.  At masses below $\sim
    1$~keV dwarf galaxies would not form~\cite{Boyarsky:08a,Gorbunov:08b}. The
    blue point would corresponds to the best-fit value from M31 if the line
    comes from DM decay. Thick errorbars are $\pm 1\sigma$ limits on the
    flux. Thin errorbars correspond to the uncertainty in the DM distribution
    in the center of M31. }
\label{fig:numsm}
\end{figure}


\textbf{Acknowledgments.} We thank D.~Malyshev for collaboration; A.~Neronov
for useful critical comments; M.~Shaposhnikov and M.~Lovell for reading the
manuscript and providing valuable comment. We also thank K.~Abazajian,
J.~Beacom, M.~Kaplinghat for their comments. The work of D.~I. was supported
by part by the the Program of Cosmic Research of the National Academy of
Sciences of Ukraine and the State Programme of Implementation of Grid
Technology in Ukraine. The work of J.F. was supported by the De Sitter program
at Leiden University with funds from NWO. This research is part of the
"Fundamentals of Science" program at Leiden University. This research has made
use of the NASA/IPAC Extragalactic Database (NED), which is operated by the
Jet Propulsion Laboratory, California Institute of Technology, under contract
with the National Aeronautics and Space Administration.

\let\jnlstyle=\rm\def\jref#1{{\jnlstyle#1}}\def\aj{\jref{AJ}}
  \def\araa{\jref{ARA\&A}} \def\apj{\jref{ApJ}\ } \def\apjl{\jref{ApJ}\ }
  \def\apjs{\jref{ApJS}} \def\ao{\jref{Appl.~Opt.}} \def\apss{\jref{Ap\&SS}}
  \def\aap{\jref{A\&A}} \def\aapr{\jref{A\&A~Rev.}} \def\aaps{\jref{A\&AS}}
  \def\azh{\jref{AZh}} \def\baas{\jref{BAAS}} \def\jrasc{\jref{JRASC}}
  \def\memras{\jref{MmRAS}} \def\mnras{\jref{MNRAS}\ }
  \def\pra{\jref{Phys.~Rev.~A}\ } \def\prb{\jref{Phys.~Rev.~B}\ }
  \def\prc{\jref{Phys.~Rev.~C}\ } \def\prd{\jref{Phys.~Rev.~D}\ }
  \def\pre{\jref{Phys.~Rev.~E}} \def\prl{\jref{Phys.~Rev.~Lett.}}
  \def\pasp{\jref{PASP}} \def\pasj{\jref{PASJ}} \def\qjras{\jref{QJRAS}}
  \def\skytel{\jref{S\&T}} \def\solphys{\jref{Sol.~Phys.}}
  \def\sovast{\jref{Soviet~Ast.}} \def\ssr{\jref{Space~Sci.~Rev.}}
  \def\zap{\jref{ZAp}} \def\nat{\jref{Nature}\ } \def\iaucirc{\jref{IAU~Circ.}}
  \def\aplett{\jref{Astrophys.~Lett.}}
  \def\apspr{\jref{Astrophys.~Space~Phys.~Res.}}
  \def\bain{\jref{Bull.~Astron.~Inst.~Netherlands}}
  \def\fcp{\jref{Fund.~Cosmic~Phys.}} \def\gca{\jref{Geochim.~Cosmochim.~Acta}}
  \def\grl{\jref{Geophys.~Res.~Lett.}} \def\jcp{\jref{J.~Chem.~Phys.}}
  \def\jgr{\jref{J.~Geophys.~Res.}}
  \def\jqsrt{\jref{J.~Quant.~Spec.~Radiat.~Transf.}}
  \def\memsai{\jref{Mem.~Soc.~Astron.~Italiana}}
  \def\nphysa{\jref{Nucl.~Phys.~A}} \def\physrep{\jref{Phys.~Rep.}}
  \def\physscr{\jref{Phys.~Scr}} \def\planss{\jref{Planet.~Space~Sci.}}
  \def\procspie{\jref{Proc.~SPIE}} \let\astap=\aap \let\apjlett=\apjl
  \let\apjsupp=\apjs \let\applopt=\ao \def\jcap{\jref{JCAP}}

\appendix
\onecolumngrid

\section{Global significance estimate}

Significances quoted in the main body of the paper (Table~I) reflect the local
significance of the signal. Since the position of the line is unknown
\textit{a priori} we need to take into account the probability of falsely
detecting a statistical fluctuation of equal or higher significance at any
position in the entire fitting range (2.0--8.0~keV). In addition, having found
a signal in the same energy bin in three separate datasets, we compute this
global significance taking into account the probability of such signals
showing in the same resolution element by chance. Given the local significance
of the signal in each dataset (based on the $\Delta\chi^2$ values and the
number of degrees of freedom), and the number of independent resolution
elements, we can determine the global significance of the combination of all
signals. The number of independent resolution elements, $N_E$, for our
datasets is about 40 (6~keV energy range divided by 150 eV --- average energy
resolution of the \xmm).

The global significance per dataset is computed from the two-sided p-value
$p_i$ (directly related to the number of $\sigma$ of the signal) by
multiplying by $N_E$ (see Table~\ref{tab:significance}). We took a ``two-sided''
p-value to take into account both positive and negative residuals.

The combined global
significance then is
\begin{equation}
\label{A1}
\frac{\prod_i p_i N_E}{N_E^{N_d-1}} = 1.1\cdot 10^{-5} 
\end{equation}
where $N_d = 3$ is the number of datasets. This corresponds to a false
detection probability for the combination dataset of \textbf{0.0011\%}.
Converted to the significance this p-value gives \textbf{$4.4\sigma$ global
  significance}.

Alternatively, we could have taken into account only probability of positive
fluctuations (so ``two-sided'' p-values in the Table~\ref{tab:significance}
should be divided by 2).  Using the same formula~\eqref{A1} we would obtain
$4.7\sigma$ global significance.

Introducing systematic uncertainties into all our datasets at the level of
$\sim 1$\%, the local significances drop by about $1\sigma$ each.

 \begin{table*}[t!]
  \centering
\begin{tabular}{l|c|c|c|c|c|c|c}
\hline
Dataset & $\Delta \chi^2$ & d.o.f. & local significance & local p-value & false detection probability & 
global significance\\
\hline
M31-oncen (MOS)& 13 & 2 & 3.18$\sigma$ & $1.5\cdot 10^{-3}$ & 0.06 & 1.89$\sigma$ \\
Perseus (MOS) & 9.1 & 2 & 2.56$\sigma$ & $1.05\cdot 10^{-2}$ & 0.42 & 0.81$\sigma$ \\
Perseus (PN) & 8 & 2 & 2.36$\sigma$ & $1.83 \cdot 10^{-2}$ & 0.73 & 0.35$\sigma$\\
\hline
All combined & & & & & $1.1 \cdot 10^{-5}$ & 4.4$\sigma$ \\
\hline
\end{tabular}
\caption{\label{tab:significance} Table of significances per dataset. 
  Quoted p-values refer to the two-sided case (one-sided p-values are half of the two-sided ones). 
  The false detection probability refers to the probability of falsely detecting a signal in 
  that dataset like the one under consideration or stronger at any energy in the range considered. 
  The global significance was converted from the false detection probability per dataset. 
  The combined false detection probability and global significance of these three datasets is 
  also given (computed from the individual detections, not from a single combined dataset).}
\end{table*}

\begin{figure}[t!]
  \centering
  \includegraphics[width=.45\textwidth]{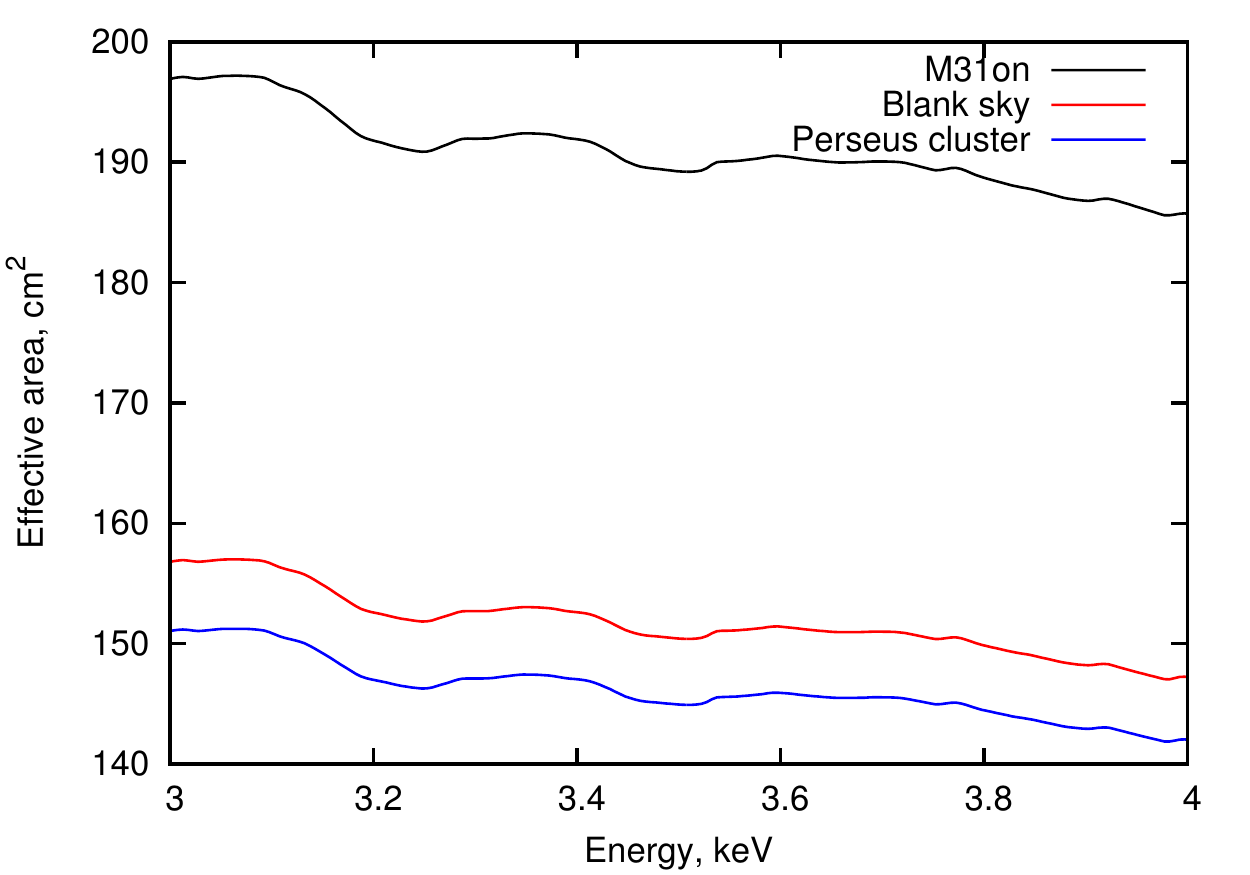}
  ~\includegraphics[width=.45\textwidth]{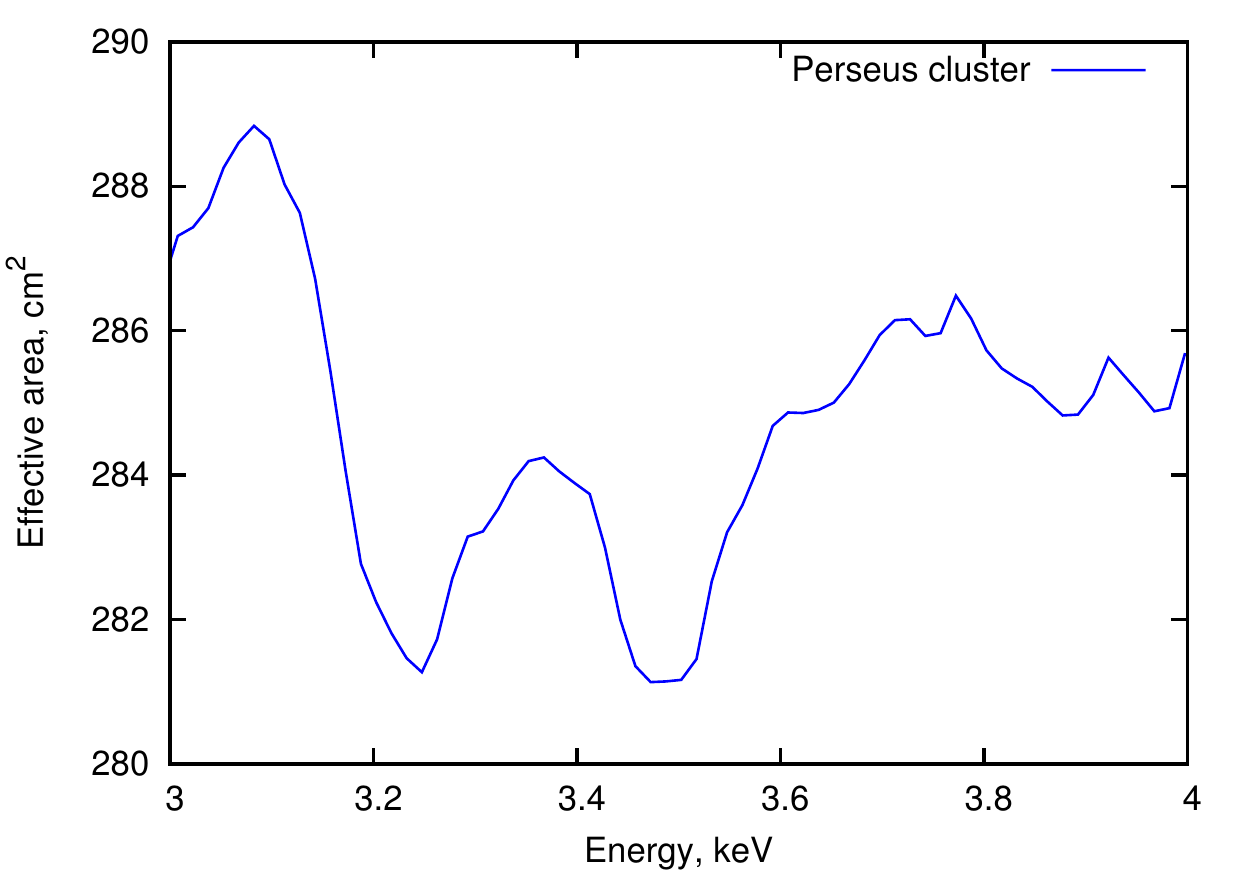}
  \caption{Exposure averaged effective area of the XMM MOS camera for the
    combination of observations of Perseus galaxy cluster, M31 and blank-sky
    (\textit{left panel}). For Perseus galaxy cluster we also show the
    exposure averaged PN camera's effective area (\textit{right panel}).}
  \label{fig:aeff}
\end{figure}

\setcounter{magicrownumbers}{0}

\begin{table*}[!p]
  \centering \begin{tabular}[c]{r|c|c|c|c|c}
    \hline
    &ObsID & Off-axis angle & Cleaned exposure & FoV  [arcmin$^2$] & F$_{in}$-F$_{out}$ \\
    &&arcmin & MOS1/MOS2 [ksec] & MOS1/MOS2 & \\
    \hline
    \rownumber & \texttt{0305690301} & 22.80 & 18.6 / 18.6 & 473.6 / 574.3 & 1.266 / 1.340 \\
    \rownumber & \texttt{0085590201} & 25.01 & 40.1 / 40.5 & 564.6 / 572.1 & 1.290 / 1.336 \\
    \rownumber & \texttt{0204720101} & 27.87 & 14.1 / 14.5 & 567.7 / 574.5 & 2.373 / 2.219 \\
    \rownumber & \texttt{0673020401} & 29.48 & 15.6 / 17.6 & 479.6 / 574.0 & 1.318 / 1.331 \\
    \rownumber & \texttt{0405410201} & 29.52 & 16.1 / 16.6 & 480.8 / 573.9 & 1.354 / 1.366 \\
    \rownumber & \texttt{0305690101} & 29.54 & 25.1 / 25.4 & 476.0 / 573.5 & 1.231 / 1.247 \\
    \rownumber & \texttt{0405410101} & 31.17 & 15.8 / 16.8 & 481.8 / 572.9 & 1.235 / 1.195 \\
    \rownumber & \texttt{0305720101} & 31.23 & 11.5 / 11.8 & 476.8 / 573.9 & 1.288 / 1.296 \\
    \rownumber & \texttt{0673020301} & 36.54 & 13.9 / 15.4 & 485.4 / 573.8 & 1.211 / 1.304 \\
    \rownumber & \texttt{0305690401} & 36.75 & 25.9 / 26.0 & 479.1 / 573.8 &
    1.158 / 1.156 \\
    \hline
    \rownumber & \texttt{0305720301} & 41.92 & 16.7 / 17.5 & 464.7 / 573.6 & 1.433 / 1.447 \\
    \rownumber & \texttt{0151560101} & 47.42 & 23.7 / 23.6 & 572.1 / 573.6 & 1.294 / 1.206 \\
    \rownumber & \texttt{0673020201} & 53.31 & 22.8 / 23.4 & 479.5 / 573.9 & 1.262 / 1.228 \\
    \rownumber & \texttt{0204720201} & 54.11 & 22.4 / 22.9 & 564.0 / 573.2 &
    1.153 / 1.195 \\
    \hline
    \rownumber & \texttt{0554500801} & 95.45 & 15.0 / 15.3 & 480.8 / 572.7 & 1.098 / 1.113 \\
    \rownumber & \texttt{0306680301} & 101.88 & 12.3 / 13.0 & 468.1 / 574.0 & 1.177 / 1.089 \\
    \hline \end{tabular} \caption{Parameters of the \xmm spectra of the Perseus
    cluster used in our analysis.  The observations are sorted by the off-axis
    angle from the center of the Perseus cluster. Two  central observations
    (ObsIDs  \texttt{0305780101} and  \texttt{0085110101}) were not included in
    the analysis to avoid modeling of the emission from the core of the
    Perseus cluster. Notice that only  these two central observations were
    used in~[14], therefore our dataset and that
    of~[14] is independent from eachother.
    The difference in FoVs between MOS1 and MOS2 cameras
    is due to the loss CCD6 in MOS1 camera. The  parameter F$_{in}$-F$_{out}$
    (last column) 
    estimates the presence of residual soft protons according to the procedure
    of~[21]. Note, however, that for the bright extended
    sources, such an estimate is not appropriate,
    see~\protect\url{http://xmm2.esac.esa.int/external/xmm_sw_cal/background/epic_scripts.shtml}
    for details). Horizontal lines shows how we group observations for
    building the surface brightness profile of the line (as shown in Fig.~2,
    right panel in the main text).}
\label{tab:perseus-xmm-observations} 
\end{table*}

\begin{table}[p!]
  \centering
  \begin{tabular}{c|c|c}
   \hline
    Range of offsets & Exposure [ksec] & Flux [$\unit{cts/sec/cm^2}$] \\
    \hline
    $23$ -- $37'$ & 400 & $13.8 \pm 3.3$\\
    $42'$ -- $54'$ & 230 & $8.3 \pm 3.4$\\
    $96'$ -- $102'$& 56 & $4.6 \pm 4.6$\\
    \hline
  \end{tabular}
  \caption{Definitions of the radial bins used for the data analysis of the Perseus cluster.}
  \label{tab:radial}
\end{table}

\begin{figure*}[t!]
  \centering
  \includegraphics[width=0.45\textwidth]{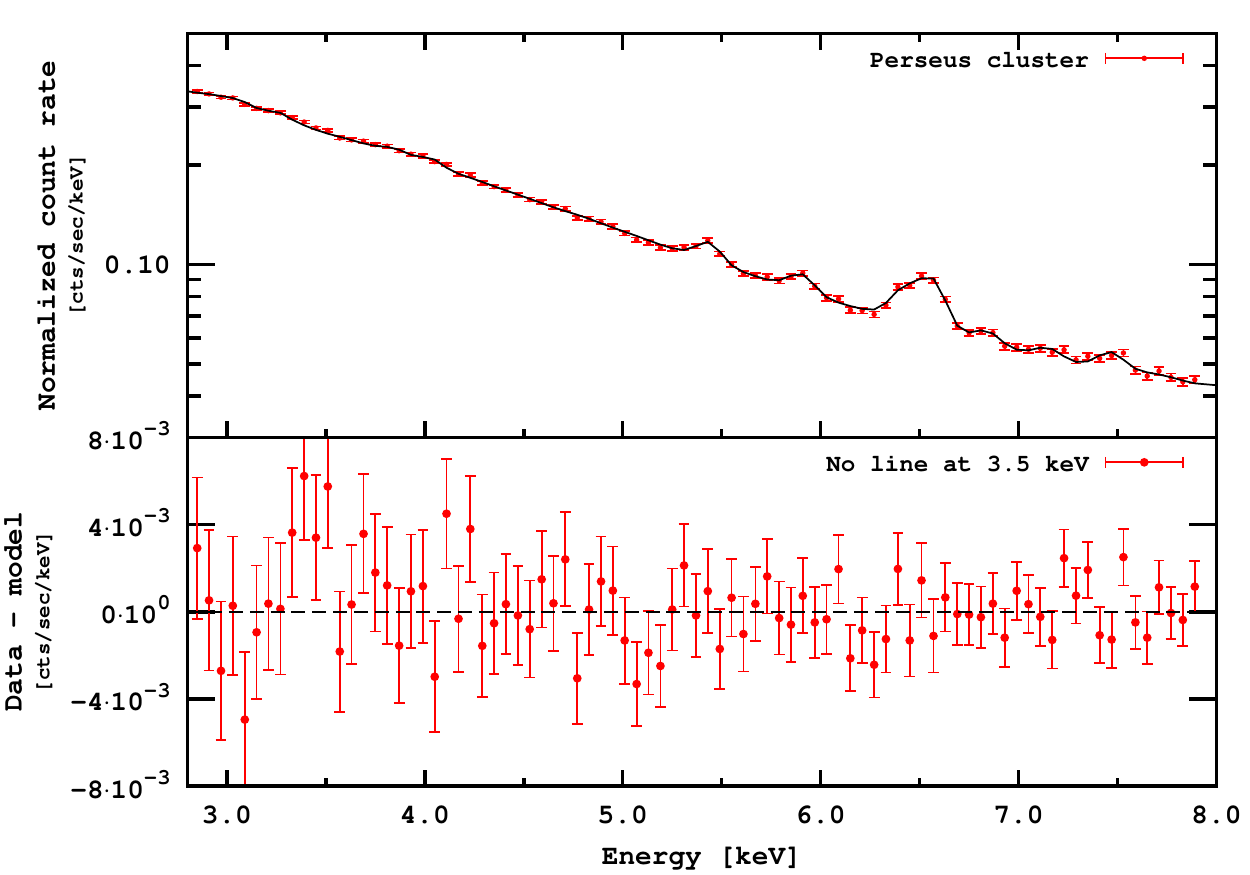}~%
  \includegraphics[width=0.45\textwidth]{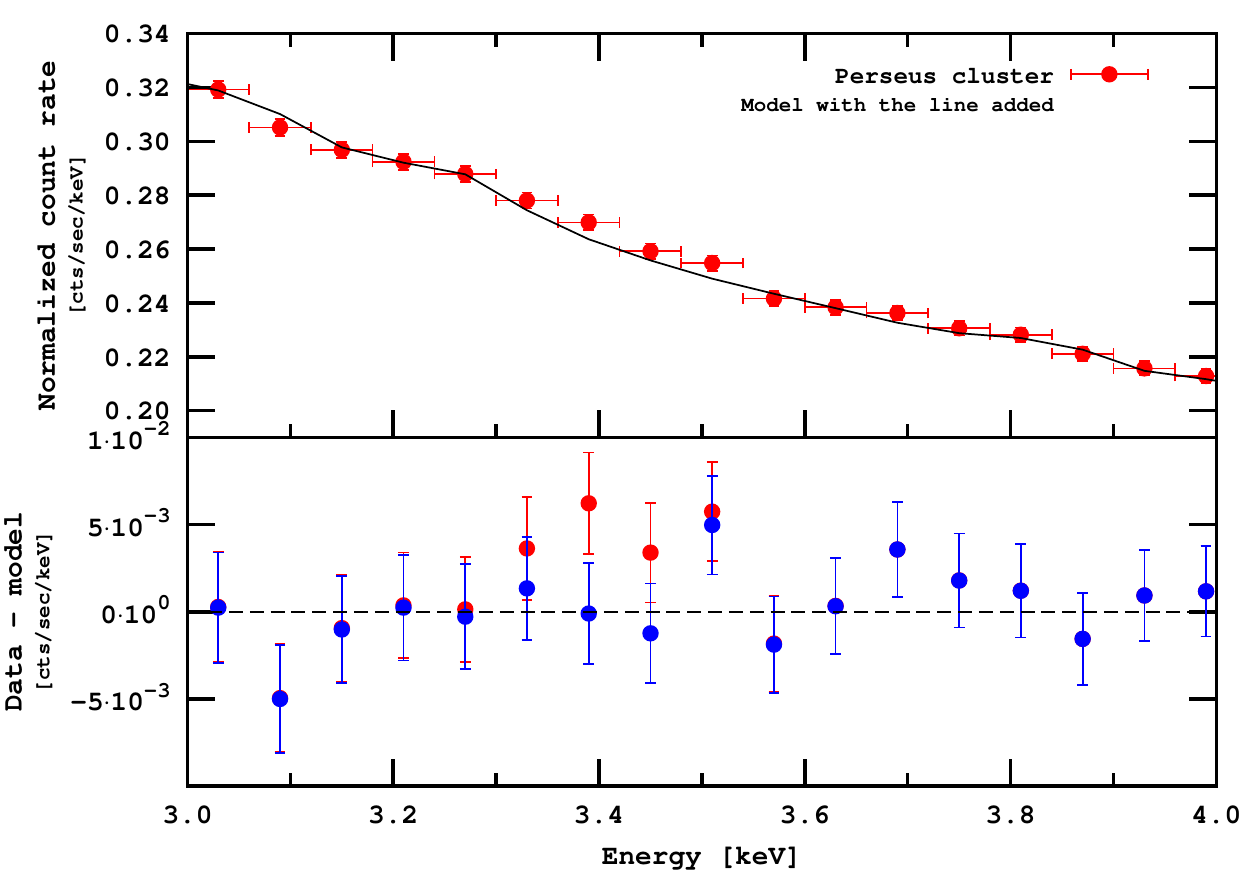}
  \caption{\textit{Left:} Folded count rate (top) and residuals (bottom) for
    the combined spectrum of 16 observations of MOS cameras (listed in the
    Tabel~\protect\ref{tab:perseus-xmm-observations}) of the 
    Perseus cluster. Statistical Y-errorbars on the top plot are smaller than
    the point size. The line around 3.5~keV is \emph{not added}, hence the
    group of positive residuals. \textit{Right}: zoom onto the line
    region. The spectrum is shown in the detector restframe, therefore the
    line is shifted left according to the Perseus redshift.}
  \label{fig:perseus}
\end{figure*}

\section{Effective area}
\label{sec:effective-area}

In this Appendix we show the effective area of the Perseus, M31 and blank-sky
datasets (Fig.~\ref{fig:aeff}).  One sees that all three datasets exhibit a
(known) wiggle at energy $E\sim 3.5$~keV in the detector frame (about $1.5\%$
deviation from the monotonic behaviour).  This kind of behavior of the effective area
is due to K-, L- and M-shell transitions of Al, Sn and Au. The SAS software
uses calibration files based on ray-tracing calculations through numerical
models of the telescope assemblies \cite{gondoin:00,mos:01,epn:01}. The
effective area curves differ between datasets mostly due to the vignetting
effect, which depends on energy and on the weighting during the data stacking.

Looking at the left panel of SOM Fig. 2 one sees that the effective area of
all MOS observations is self-similar. The variation in shape between three
datasets in the energy range 3.4-3.6 keV is less than 0.1\% and less than 0.4\% 
in the 3-4 keV range.  If the line is due to an unmodeled wiggle, this
would mean that a 10 times larger unmodeled feature (line is 3-4\% of the
continuum level) is present in the datasets of M31 and Perseus, but \emph{not}
in the blank sky. As all datasets are combinations of observations taken over
long period of lifetime of the XMM, the existence of such a feature is
difficult to imagine.

Notice that if this wiggle would be
the cause of the signal, reported in this paper, it would fail to explain why
the redshift of the line in the Perseus cluster is correctly detected (at
energies $3.5/1+z=3.4$~keV the effective area has a local maximum, rather
than minimum). It would also fail to explain the detection of the line in the
combined dataset of 70 clusters at different redshifts, presented
in~\cite{Bulbul:14}. 

Additionally, if the feature is due to an \emph{unmodelled} wiggle in the
effective area, its flux in each dataset should be proportional to the
continuum brightness.  Comparing the M31 and blank-sky datasets we see that
the count rate at energies of interest is 4 times larger for M31, so that the
blank-sky dataset would have exhibited a 4 times smaller (instrumental)
feature with a flux $\sim \unit[1.2 \times 10^{-6}]{cts/sec/cm^2}$, were it
due to a wiggle in the effective area.  Notice that the exposure for the blank
sky is 16 times larger and such a line would have been resolved with
sufficient statistical significance. The upper (non-detection) limit from the
blank-sky dataset is $\sim 2$ lower ($\unit[0.7 \times
10^{-6}]{cts/sec/cm^2}$).

\section{Flare removal}
\label{sec:flare-removal}

\begin{figure}[!t]
  \centering
  \includegraphics[width=.5\textwidth]{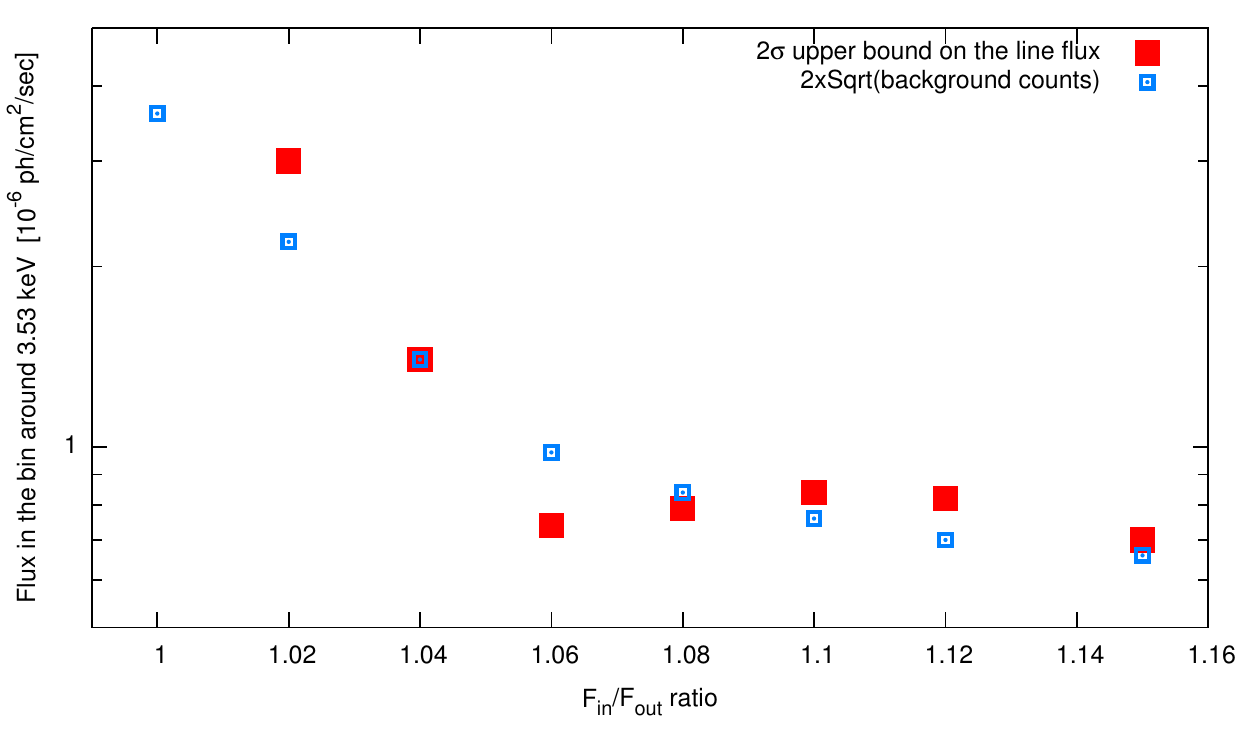}
  \caption{The dependence of the $2\sigma$ upper bound on the flux in the
    blanksky dataset on the imposed $F_{in}-F_{out}$ criterion. The
    statistical error on this parameter is about $5\%$. The bound on the flux
    remains at the quoted level until we start to lose significant fraction of
    observations for $F_{in}-F_{out} < 1.06$. Blue squares are defined as
    $2\times \sqrt{N_\text{bg}}$ where $N_\text{bg}$ is the number of
    background counts in the energy bin, equal to spectral resolution. The
    difference between blue and red squares appears because spectral modeling
    trakes into account also the line shape.}
  \label{fig:FinFout}
\end{figure}

In this Section we investigate how sensitive the derived bounds are to the
imposed $F_{in}-F_{out}$ cut. To this end we have imposed a number of
different cuts in $F_{in}-F_{out}$ and rederived the $2\sigma$ upper bound in
the blank sky dataset. We see (Fig.~\ref{fig:FinFout}) that the bound derived
in the paper does not really change until we start to impose very stringent
cuts $F_{in}-F_{out} < 1.06$, which starts to drastically reduce the statistics
(clean exposure) as the blue squares in Fig.~\ref{fig:FinFout} demonstrate).

\begin{figure}[!t]
  \centering
  \includegraphics[width=0.5\textwidth]{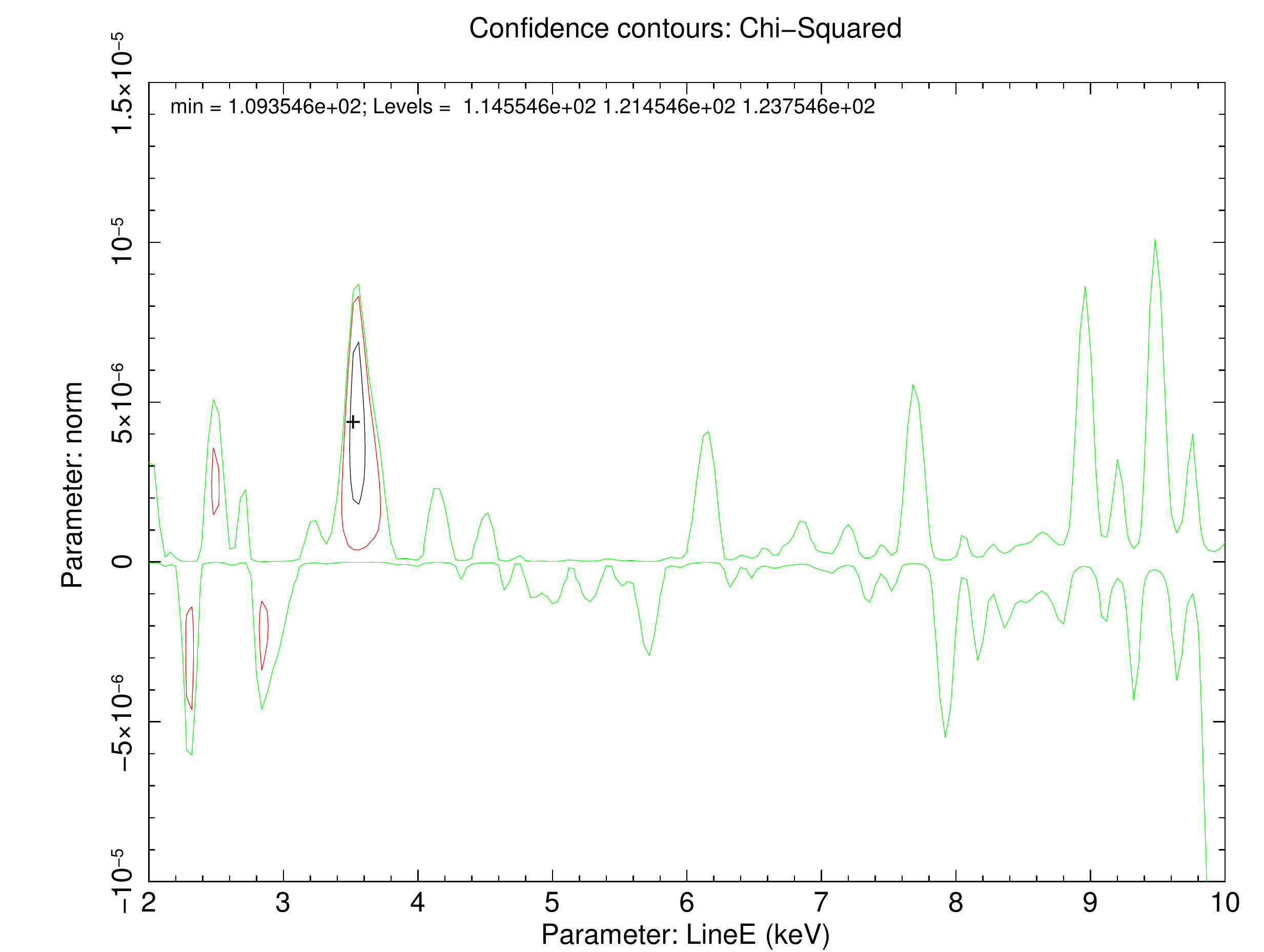}
  \caption{Structure of the residuals (both positive and negative) around the
    best fit model for M31 central observation. Red contours show residuals
    that are above $1\sigma$. Black contour shows more than $3\sigma$ residual
    ($3.53$~keV line). The other residuals are below $1\sigma$.}
  \label{fig:resM31}
\end{figure}

\begin{table}[!p]
\centering \relsize{-0.6}
\begin{tabular}[c]{r|c|c|c|c|c}
\hline
 & ObsID & Off-axis angle & Cleaned exposure & FoV [arcmin$^2$] & F$_{in}$-F$_{out}$ \\
   && arcmin & MOS1/MOS2 [ksec] & MOS1/MOS2 & \\
   \hline
   \rownumber & \texttt{0405320501} & 0.02 & 12.3/13.6 & 480.6/573.2 & 1.132/1.039 \\
   \rownumber & \texttt{0405320701} & 0.02 & 14.8/14.9 & 480.7/572.8 & 1.046/1.057 \\
   \rownumber & \texttt{0405320801} & 0.02 & 13.1/13.1 & 488.2/573.0 & 1.160/1.117 \\
   \rownumber & \texttt{0405320901} & 0.02 & 15.5/15.6 & 488.0/574.3 & 1.099/1.065 \\
   \rownumber & \texttt{0505720201} & 0.02 & 25.2/26.2 & 485.6/572.1 & 1.079/1.057 \\
   \rownumber & \texttt{0505720301} & 0.02 & 25.4/24.3 & 486.0/573.9 & 1.129/1.105 \\
   \rownumber & \texttt{0505720401} & 0.02 & 19.9/20.2 & 488.6/573.1 & 1.113/1.108 \\
   \rownumber & \texttt{0505720501} & 0.02 & 12.9/13.9 & 480.3/574.1 & 1.151/1.064 \\
   \rownumber & \texttt{0505720601} & 0.02 & 20.2/20.4 & 488.3/571.4 & 1.085/1.108 \\
   \rownumber & \texttt{0551690201} & 0.02 & 20.5/20.3 & 486.5/574.2 & 1.099/1.072 \\
   \rownumber & \texttt{0551690301} & 0.02 & 19.7/19.4 & 479.3/573.0 & 1.109/1.117 \\
   \rownumber & \texttt{0551690501} & 0.02 & 16.9/18.4 & 486.3/573.2 & 1.095/1.109 \\
   \rownumber & \texttt{0600660201} & 0.02 & 17.4/17.5 & 487.0/572.9 & 1.080/1.041 \\
   \rownumber & \texttt{0600660301} & 0.02 & 16.1/16.1 & 488.6/572.0 & 1.054/1.041 \\
   \rownumber & \texttt{0600660401} & 0.02 & 15.0/15.5 & 479.9/573.1 & 1.078/1.072 \\
   \rownumber & \texttt{0600660501} & 0.02 & 13.5/14.3 & 488.2/573.4 & 1.079/1.083 \\
   \rownumber & \texttt{0600660601} & 0.02 & 15.2/15.1 & 481.8/573.6 & 1.073/1.041 \\
   \rownumber & \texttt{0650560201} & 0.02 & 21.0/21.3 & 488.1/573.3 & 1.198/1.140 \\
   \rownumber & \texttt{0650560301} & 0.02 & 26.9/29.0 & 487.9/572.6 & 1.082/1.095 \\
   \rownumber & \texttt{0650560401} & 0.02 & 12.4/13.5 & 488.0/573.1 & 1.157/1.069 \\
   \rownumber & \texttt{0650560501} & 0.02 & 15.8/21.6 & 487.8/573.4 & 1.162/1.114 \\
   \rownumber & \texttt{0650560601} & 0.02 & 20.8/21.5 & 487.5/572.2 & 1.085/1.068 \\
   \rownumber & \texttt{0674210201} & 0.02 & 19.6/19.6 & 478.6/573.3 & 1.094/1.083 \\
   \rownumber & \texttt{0674210301} & 0.02 & 14.9/15.0 & 488.1/573.6 & 1.052/1.043 \\
   \rownumber & \texttt{0674210401} & 0.02 & 17.9/18.1 & 485.7/572.7 & 1.071/1.081 \\
   \rownumber & \texttt{0674210501} & 0.02 & 16.2/16.3 & 488.8/573.5 & 1.192/1.139 \\
   \rownumber & \texttt{0202230201} & 1.44 & 18.3/18.4 & 567.1/572.8 & 1.089/1.108 \\
   \rownumber & \texttt{0202230401} & 1.44 & 17.0/17.1 & 566.5/573.6 & 1.118/1.109 \\
   \rownumber & \texttt{0202230501} & 1.44 & 9.2/9.4 & 568.1/574.1 & 1.048/1.129 \\
\hline
   \rownumber & \texttt{0402560201} & 23.71 & 16.0/16.6 & 478.7/574.0 & 1.096/1.095 \\
   \rownumber & \texttt{0505760201} & 23.71 & 35.2/38.6 & 476.6/571.6 & 1.065/1.058 \\
   \rownumber & \texttt{0511380201} & 23.71 & 15.3/15.4 & 485.0/572.7 & 1.126/1.047 \\
   \rownumber & \texttt{0511380601} & 23.71 & 14.8/17.2 & 485.4/573.1 & 1.041/1.074 \\
   \rownumber & \texttt{0402560901} & 24.18 & 42.4/42.9 & 475.0/572.8 & 1.118/1.071 \\
   \rownumber & \texttt{0672130101} & 24.24 & 73.0/78.6 & 473.1/572.8 & 1.088/1.064 \\
   \rownumber & \texttt{0672130501} & 24.24 & 22.7/25.4 & 477.0/574.8 & 1.097/1.110 \\
   \rownumber & \texttt{0672130601} & 24.24 & 67.8/67.3 & 471.8/571.4 & 1.115/1.101 \\
   \rownumber & \texttt{0672130701} & 24.24 & 70.7/74.3 & 484.8/573.5 & 1.076/1.052 \\
   \rownumber & \texttt{0410582001} & 26.29 & 13.2/13.9 & 485.4/575.0 & 1.073/1.030 \\
   \rownumber & \texttt{0402561001} & 28.81 & 48.0/49.4 & 478.4/572.5 & 1.084/1.042 \\
   \rownumber & \texttt{0402560301} & 30.34 & 43.9/45.7 & 474.6/573.1 & 1.037/1.027 \\
   \rownumber & \texttt{0505760301} & 39.55 & 41.0/41.3 & 485.0/570.8 & 1.022/1.022 \\
   \rownumber & \texttt{0402561101} & 39.56 & 44.8/44.8 & 478.7/571.4 & 1.121/1.067 \\
   \rownumber & \texttt{0404060201} & 42.94 & 19.1/19.1 & 480.7/573.7 & 0.993/1.045 \\
   \rownumber & \texttt{0402561201} & 47.37 & 38.1/39.2 & 478.5/573.3 & 1.077/1.034 \\
   \rownumber & \texttt{0402560501} & 49.06 & 48.8/50.6 & 487.2/572.9 & 1.102/1.079 \\
   \rownumber & \texttt{0511380301} & 49.06 & 31.5/31.0 & 482.0/572.3 & 1.105/1.082 \\
   \rownumber & \texttt{0151580401} & 50.89 & 12.3/12.3 & 567.2/574.1 & 1.131/1.020 \\
   \rownumber & \texttt{0109270301} & 55.81 & 25.5/25.0 & 562.6/571.6 & 1.110/1.106 \\
   \hline
\end{tabular}
\caption{Parameters of the \xmm\ spectra of M31 used in our analysis.  The
  significant difference in FoVs between MOS1 and MOS2 cameras is due to the
  loss CCD6 in MOS1 camera. The F$_{in}$-F$_{out}$ parameter estimating the
  presence of residual soft protons according to procedure of~[21]
  is also shown. First 29 observations are the dataset M31 ON, the remaining
  20 -- the dataset M31 OFF.}
\label{tab:m31-xmm-observations} 
\end{table}
   
  \end{document}